\documentclass[%
 reprint,
showpacs,preprintnumbers,
 amsmath,amssymb,
 aps,
]{revtex4-1}

\usepackage[dvipdfmx]{graphicx}
\usepackage{epstopdf}
\usepackage{dcolumn}
\usepackage{bm}
\usepackage{color}
\usepackage{ulem}

\begin{document}

\title{
Nonequilibrium-relaxation approach to quantum phase transitions:\\
Nontrivial critical relaxation in cluster-update quantum Monte Carlo
}

\author{Yoshihiko Nonomura}
\email{nonomura.yoshihiko@nims.go.jp}
\affiliation{International Center for Materials Nanoarchitectonics, 
National Institute for Materials Science, Tsukuba, Ibaraki 305-0044, Japan} 

\author{Yusuke Tomita}
\email{ytomita@shibaura-it.ac.jp}
\affiliation{College of Engineering, Shibaura Institute of Technology, 
Saitama 337-8570, Japan} 

\date{\today}

\begin{abstract}
Although the nonequilibrium relaxation (NER) method has been widely used 
in Monte Carlo studies on phase transitions in classical spin systems, such 
studies have been quite limited in quantum phase transitions. The reason 
is that relaxation process based on cluster-update quantum Monte Carlo 
(QMC) algorithms, which are now standards in Monte Carlo studies on 
quantum systems, has been considered ``too fast" for such analyses. 
Recently the present authors revealed that the NER process in classical 
spin systems based on cluster-update algorithms is characterized by 
the stretched-exponential critical relaxation, rather than the conventional 
power-law one in local-update algorithms. In the present article we 
show that this is also the case in quantum phase transitions analyzed 
with the cluster-update QMC, and that advantages of NER analyses 
are available. As the simplest example of isotropic quantum spin models 
which exhibit quantum phase transitions, we investigate the N\'eel-dimer 
quantum phase transition in the two-dimensional $S=1/2$ columnar-dimerized 
antiferromagnetic Heisenberg model with the continuous-time loop algorithm.
\end{abstract}


\maketitle

{\it Introduction}. 
Dynamical Monte Carlo (MC) simulations have been widely utilized 
in statistical-mechanical studies on classical spin systems, where 
the Boltzmann weight for a MC flip is determined locally. 
In quantum spin systems, the Boltzmann weight 
cannot be determined locally unless diagonalizing 
the Hamiltonian matrix in principle, and treatable system sizes are 
strictly reduced. This difficulty was overcome by introducing the 
Suzuki-Trotter decomposition~\cite{Trotter59,Suzuki76}, where 
noncommutable exponential operators are approximately divided 
into $n$ Trotter layers, and the original system is reproduced 
in the $n \to \infty$ limit. This procedure can be regarded as 
introduction of an extra imaginary-time axis.

In the original formulation of quantum Monte Carlo (QMC)~\cite{Suzuki76}, 
the $n \to \infty$ limit is taken by numerical extrapolation with several 
finite-$n$ systems, and nontrivial global-update processes should 
be introduced additionally by hands in order to keep ergodicity. 
Such a complicated procedure was simplified by the continuous-time 
QMC algorithm~\cite{Beard96}, where the $n \to \infty$ limit is taken 
as a part of formulation without numerical extrapolation. This QMC 
algorithm should be coupled with cluster-update QMC algorithms such as 
the loop algorithm~\cite{Evertz93}, worm algorithm~\cite{Prokovev98}, 
or directed-loop algorithm~\cite{Syljuasen02}, in which ergodicity is 
ensured within the formulation.

The nonequilibrium relaxation (NER) method~\cite{NERrev} is one of the MC 
approaches to investigate phase transitions against the critical slowing down. 
In contrast to other approaches such as the cluster algorithms~\cite{SW,Wolff} 
and extended ensemble methods~\cite{Berg,Hukushima,WL}, 
the critical slowing down is not avoided but rather utilized in this scheme 
for the evaluation of critical phenomena. The critical point is estimated 
from the power-law behavior of physical quantities expected there, 
and the critical exponents are evaluated from the exponents of 
such relaxation behaviors. This relaxation process is usually 
terminated much earlier than arrival at equilibrium. 
In order to avoid artifacts originating from biased initial 
states, completely-ordered or completely-disordered states 
are usually chosen as initial ones for the relaxation process.

In equilibrium MC simulations, thermal averaging is taken 
during long-time measurement after discarding equilibration data. 
In NER calculations, such averaging is replaced by that for independent 
random-number sequences (RNS), and all the numerical data are utilized. 

Several attempts to generalize the NER scheme to above-mentioned 
modern QMC algorithms had not been successful so far, and 
cluster-update QMC algorithms had been considered ``too fast" 
for NER. Succeeded examples of the application of NER to QMC 
were based on a world-line local-update QMC algorithm with 
finite Trotter layers~\cite{Nonomura98} or a continuous-time QMC 
algorithm only along the imaginary-time axis~\cite{Nakamura03}, 
both of which simply slowed down relaxation to fit for power-law NER.

Recently the present authors numerically found that the early-time 
nonequilibrium critical relaxation in cluster algorithms is described by the 
stretched-exponential simulation-time dependence, not by the power-law one 
in various classical spin systems~\cite{Nonomura14,Nonomura15,Nonomura16}. 
NER framework for cluster algorithms can be constructed on the basis of this 
relaxation formula. Quite recently the present authors derived this relaxation formula 
phenomenologically in the Ising models in the Swendsen-Wang algorithm~\cite{Tomita18}.

In the present article, we propose an NER analysis of quantum 
phase transitions based on the continuous-time loop algorithm. 
Although this algorithm is based on cluster update, it is not trivial 
if the stretched-exponential critical relaxation is observed in the 
present one-dimensional loop clusters, which is different from 
bulky ones in cluster algorithms in classical spin systems. 

\begin{figure}[b]
\includegraphics[width=50mm]{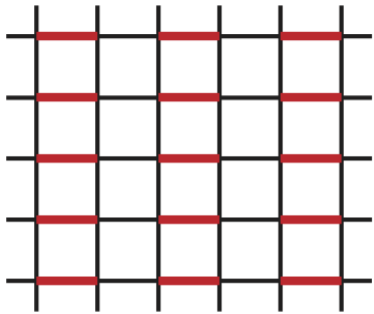}
\caption{
Schematic figure of the columnar dimer pattern. 
Broad red lines stand for the dimerized bonds.
}
\label{dimfig}
\end{figure}
\smallskip
\par
{\it Model and method}. 
In the present article we investigate the two-dimensional $S=1/2$ 
columnar-dimerized antiferromagnetic Heisenberg model on a 
square lattice,
\begin{eqnarray}
\label{dimham}
{\cal H} &=& \sum_{\langle ij \rangle \in {\rm n.n.}} 
      \hspace{-2mm} J_{ij} \vec{S}_{i} \cdot \vec{S}_{j},\ S=1/2,\\
J_{ij} &=& \left\{
\begin{tabular}{rc}
$(1+\delta)J$ & {\rm \ \ on dimerized bonds ($\delta \geq 0$),}\\
$J$ & {\rm elsewhere},
\end{tabular}
\right.
\end{eqnarray}
with the nearest-neighbor interaction and the columnar dimer pattern 
shown in Fig.\ \ref{dimfig}. 
For $\delta=0$ (the uniform case), this model has the N\'eel order 
(reduced due to quantum fluctuations) in the ground state. For large enough 
$\delta$, singlet pairs on the dimerized bonds wipe out the N\'eel order. 
There exists a critical point $\delta_{\rm c}$ between the two cases, and 
this quantum phase transition is of conventional second order, because 
the dimerization does not accompany spontaneous symmetry breaking. 

This is the simplest isotropic quantum spin model with quantum 
phase transition without frustration. This model has been intensively 
studied with QMC~\cite{Matsumoto01,Wenzel09,Yasuda13}, and  
its universality class has been considered to be the same 
as that of the three-dimensional (3D) classical Heisenberg model. 
Here we analyze the early-time relaxation behavior of this model 
with the continuous-time loop algorithm. In the NER analysis, 
choice of the initial state is crucial. 
In the 3D classical Heisenberg model, we found that 
the ordering process from the perfectly-disordered state gives much 
smaller deviation from the stretched-exponential critical relaxation than 
the decaying process from the perfectly-ordered state~\cite{Nonomura16}.

In classical spin systems, the perfectly-disordered state can 
be one of the states for $T \to \infty$. However, when NER 
calculations are started from such states in quantum phase 
transitions, they quickly converge to a classical ordered state 
corresponding to the basis used in the MC algorithms, and 
then converge to equilibrium at the quantum critical point. 
In the present study, we utilize the continuous-time loop 
algorithm formulated with the Ising basis, and measure 
physical quantities based on the classical N\'eel order. 
MC time evolution is based on the standard loop algorithm; 
loop clusters are generated in the whole system and each 
cluster is flipped with $50$\% probability, similarly to the 
Swendsen-Wang algorithm in classical spin systems~\cite{SW}. 
Then, we start from the isolated dimer state, in which 
only singlet pairs are on the dimerized bonds, 
and it becomes the ground state in the $\delta \to \infty$ limit. 
Since the parameter $\delta$ plays a role of temperature 
in this quantum phase transition, this process corresponds 
to NER from a perfectly-disordered state.

In the present study, we analyze the absolute value of the N\'eel order 
measured on the initial Trotter layer, 
\begin{equation}
\label{Neel}
m_{\rm N} \equiv \frac{1}{N}\sum_{i}(-1)^{i}S_{i}^{z},
\end{equation}
with abbreviations $i \equiv (i_{x},i_{y})$ and $(-1)^{i} \equiv (-1)^{i_{x}+i_{y}}$. 
In QMC simulations of antiferromagnets on bipartite lattices, 
the original Hamiltonian (\ref{dimham}) is transformed to 
\begin{equation}
{\cal H} = \sum_{\langle ij \rangle \in {\rm n.n.}} 
\hspace{-2mm} J_{ij} \left( -S_{i}^{x} S_{j}^{x} -S_{i}^{y} S_{j}^{y} + S_{i}^{z} S_{j}^{z} \right),\ S=1/2,
\end{equation}
with the spin rotation $S_{i}^{x} \to -S_{i}^{x}$, $S_{i}^{y} \to -S_{i}^{y}$, 
$S_{i}^{z} \to S_{i}^{z}$ on one of the sublattices in order to refrain from the 
negative-sign problem. Then, the singlet state on each dimerized bond is transformed 
from $(\vert \hspace{-1mm} \uparrow\downarrow \rangle - \vert \hspace{-1mm} \downarrow\uparrow \rangle)/\sqrt{2}$ 
to $(\vert \hspace{-1mm} \uparrow\downarrow \rangle + \vert \hspace{-1mm} \downarrow\uparrow \rangle)/\sqrt{2}$
in actual simulations, and typical states consisting of the isolated dimer state 
in the transformed system with the Ising basis are constructed as follows:
\begin{enumerate}
\item Consider the system with $J_{ij}=(1 + \delta) J$ on dimerized bonds and $J_{ij}=0$ elsewhere.
\item Assign the basis $\vert \hspace{-1.2mm} \uparrow \downarrow \rangle$ 
or $\vert \hspace{-1.2mm} \downarrow \uparrow \rangle$ on each dimerized bond 
randomly but to keep the N\'eel order (\ref{Neel}) vanishing 
on the initial Trotter layer, on which physical quantities are measured. 
(When the RNS average is taken, this initial configuration is also changed.) 
\item Insert gates along the imaginary-time direction~\cite{QMCrev} on each dimerized bond with 
the probability corresponding to $J_{ij}=(1+\delta)J$. The number of gates should be even on each 
dimerized bond in order to satisfy the periodic boundary condition along the imaginary-time direction.
\item Flip the basis $\vert \hspace{-1mm} \uparrow \downarrow \rangle$ to 
$\vert \hspace{-1mm} \downarrow \uparrow \rangle$ and vice versa at each gate. 
(When the steps 3 and 4 are skipped, we have one of 
the classical perfectly-disordered states.) 
\end{enumerate}

When the stretched-exponential critical relaxation holds, 
early-time behavior of the absolute value of the N\'eel order 
at the quantum critical point $\delta_{\rm c}$ is given by 
\begin{equation}
\label{se-sc}
\langle |m_{\rm N}(t,L)| \rangle 
\sim L^{-d/2} \exp \left( + c_{m} t ^{\sigma} \right),
\end{equation}
with the RNS average $\langle \cdots \rangle$, 
simulation time $t$, linear size $L_{x}=L_{y}=L$, 
spatial dimension $d=2$ ({this size dependence originates from 
the normalized random-walk growth of spin clusters, and here we 
do not take the summation along the imaginary-time direction), 
quantity-dependent coefficient $c_{m}$ and relaxation exponent 
$\sigma$ ($0<\sigma<1$) common in all the physical quantities. 
Combining this formula with the equilibrium finite-size scaling relation, 
namely $\langle |m_{\rm N}(t=\infty,L)| \rangle \sim L^{-\beta/\nu}$, 
we arrive at the following nonequilibrium-to-equilibrium scaling relation, 
\begin{equation}
\label{mNneqsc}
\langle |m_{\rm N}(t,L)| \rangle L^{\beta/\nu}
\sim f_{m}( c_{m} t^{\sigma}-\ln L^{d/2-\beta/\nu} ),
\end{equation}
with a quantity-dependent scaling function $f_{m}$~\cite{Nonomura14,Nonomura16}.
\begin{figure}
\includegraphics[width=88mm]{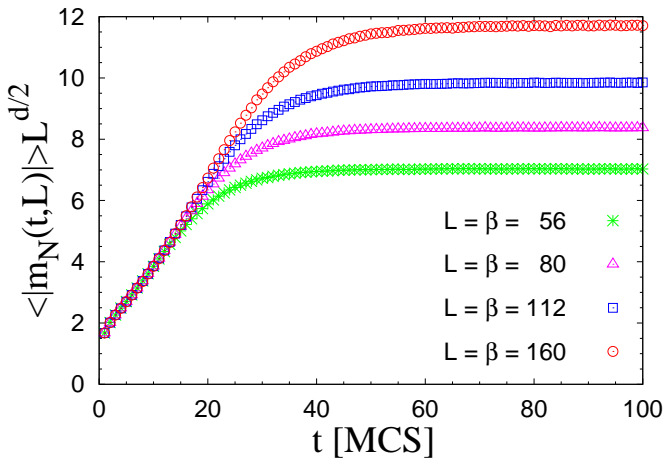}
\caption{Simulation-time dependence of the absolute value of 
the N\'eel order at $\delta=0.90947$ for $L=\beta=56$ (green stars), 
$80$ (pink triangles), $112$ (blue squares) and $160$ (red circles). 
}
\label{rawfig}
\end{figure}
\smallskip
\par
{\it Numerical results}. 
We analyze the absolute value of the N\'eel order of the model 
(\ref{dimham}) and evaluate the quantum critical point 
$\delta_{\rm c}$, critical exponent $\beta/\nu$ and 
relaxation exponent $\sigma$. As an example, this quantity multiplied 
with $L^{d/2}$ at $\delta=0.90947$ (as will be shown below, the most 
probable value of $\delta_{\rm c}$) is plotted versus simulation time in 
Fig.~\ref{rawfig} for $L=56$ ($2.56 \times 10^{6}$ RNS are averaged), 
$80$ ($2.56 \times 10^{6}$ RNS), $112$ ($1.28 \times 10^{6}$ RNS) 
and $160$ ($0.64 \times 10^{6}$ RNS), 
where the system size along the imaginary-time axis is taken the same as 
those along the real axes, namely $\beta \equiv 1/(k_{\rm B}T) = L$. 
As expected, this quantity is scaled well with Eq.~(\ref{se-sc}) initially, and 
tends to be away from this equation and to saturate as $t$ increases.

\begin{figure}
\includegraphics[width=88mm]{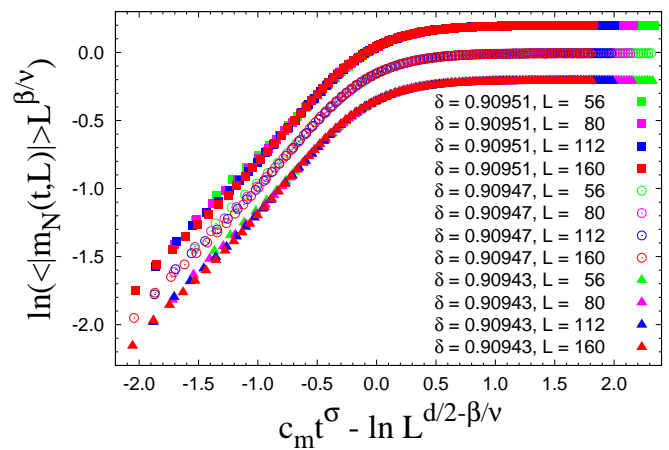}
\caption{
Nonequilibrium-to-equilibrium scaling plot of the absolute 
value of the N\'eel order at $\delta=0.90947$ (circles) 
for $L=\beta=56$ (green symbols), $80$ (pink symbols), 
$112$ (blue symbols) and $160$ (red symbols) with 
$\beta/\nu=0.514(1)$, $\sigma=0.502(8)$, $c_{m}=0.423(14)$.
The data at $\delta=0.90951$ (full squares) and $0.90943$ 
(full triangles) for the same sizes are also scaled with 
$\beta/\nu=0.515(1)$, $\sigma=0.501(8)$, $c_{m}=0.427(16)$ and 
$\beta/\nu=0.513(1)$, $\sigma=0.504(7)$, $c_{m}=0.421(13)$,  respectively.
}
\label{fitfig}
\end{figure}
The data in Fig.~\ref{rawfig} are scaled with Eq.~(\ref{mNneqsc}) 
in Fig.~\ref{fitfig} with $\beta/\nu=0.514(1)$, $\sigma=0.502(8)$ and 
$c_{m}=0.423(14)$. The exponent $\beta/\nu$ is evaluated from the 
scaling behavior in the vicinity of equilibrium, which little depends on 
the values of $\sigma$ and $c_{m}$. Details of this evaluation will 
be explained later together with the estimation of $\delta_{\rm c}$. 
The values of $\sigma$ and $c_{m}$ are evaluated so as to minimize 
the mutual residue of the data with the estimate of $\beta/\nu$. 
In Ref.~\cite{Nonomura16}, we also used the criterion that the 
early-time behavior is described by the stretched-exponential 
relaxation formula (\ref{se-sc}), and the slope of the scaled data should 
be unity in the region $ c_{m} t^{\sigma}-\ln L^{d/2-\beta/\nu} \lesssim -0.5$ 
in Fig.\ \ref{fitfig}. Here we do not use this criterion, because validity of 
this formula cannot be fully justified~\cite{comment}.
\begin{figure}
\includegraphics[width=88mm]{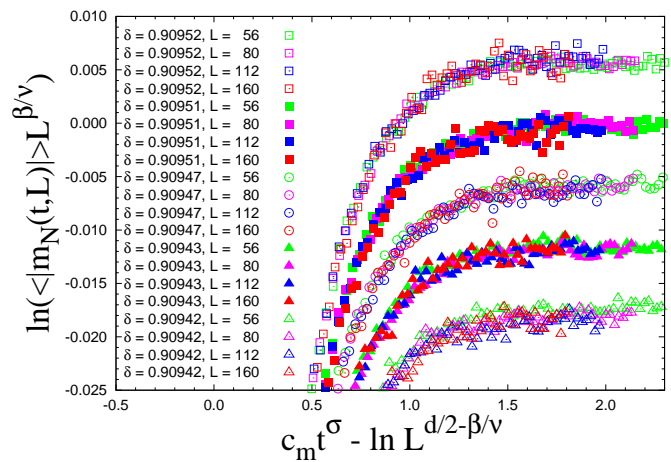}
\caption{Nonequilibrium-to-equilibrium scaling plot of the 
absolute value of the N\'eel order in the vicinity of equilibrium 
at $\delta=0.90952$ (open squares with $\beta/\nu=0.516$), 
$0.90951$ (full squares with $\beta/\nu=0.515$), 
$0.90947$ (open circles with $\beta/\nu=0.514$), $0.90943$ 
(full triangles with $\beta/\nu=0.513$) and $0.90942$ 
(open triangles with $\beta/\nu=0.512$) for $L=\beta=56$ 
(green symbols), $80$ (pink symbols), $112$ (blue symbols) 
and $160$ (red symbols) with $\sigma=0.502$ and 
$c_{m}=0.423$ used in Fig.~\ref{fitfig} at $\delta=0.90947$ 
for all the data in this figure.
}
\label{expfig}
\end{figure}

In order to evaluate the quantum critical point $\delta_{\rm c}$, 
the data in the vicinity of equilibrium for $L=56$ (green symbols), 
$80$ (pink symbols), $112$ (blue symbols) and $160$ (red symbols) 
at $\delta=0.90952$ (open squares with $\beta/\nu=0.516$), $0.90951$ 
(full squares with $\beta/\nu=0.515$), $0.90947$ (open circles with 
$\beta/\nu=0.514$), $0.90943$ (full triangles with $\beta/\nu=0.513$) 
and $0.90942$ (open triangles with $\beta/\nu=0.512$) are plotted in 
Fig.~\ref{expfig} with $\sigma=0.502$ and $c_{m}=0.423$ estimated 
at $\delta=0.90947$ in Fig.~\ref{fitfig}. 
The data except for $\delta=0.90947$ are slightly shifted upwards or downwards 
for clear visualization. Although the data at $\delta=0.90951$ and $0.90943$ 
are still scaled on a single curve with $\beta=0.515(1)$, $\sigma=0.501(8)$, 
$c_{m}=0.427(16)$ and $\beta=0.513(1)$, $\sigma=0.504(7)$, $c_{m}=0.421(13)$, 
respectively (see Fig.~\ref{fitfig}; these data are shifted upwards or downwards for clear 
visualization), such a behavior is not observed at $\delta=0.90952$ and $0.90942$. 
Here the data for $L=80$ and $112$ are tuned to be scaled on a 
single curve. Then, deviation of the data for $L=56$ and $80$ 
and that for $L=112$ and $160$ are in opposite directions, 
and all the data cannot be scaled anymore. 
In order to include all the results, our final estimates are given by 
\begin{eqnarray}
&& \delta_{\rm c}=0.90947(5),\\
&& \beta/\nu=0.514(2),\ \sigma=0.502(9),\ c_{m}=0.426(17).
\end{eqnarray}
These estimates are consistent with the previous ones for the same 
model based on the equilibrium QMC simulations~\cite{Yasuda13}, 
$\delta_{\rm c}=0.90947(3)$ and $\beta/\nu=0.513(9)$, 
and with ours for the 3D classical Heisenberg model 
based on the cluster NER~\cite{Nonomura16}, 
$\beta/\nu=0.515(5)$ and $\sigma \approx 1/2$. 
Although the evaluation of $\delta_{\rm c}$ based on the deviation of data in 
Fig.~\ref{expfig} is rather subtle, that of $\beta/\nu$ based on the scaling formula 
(\ref{mNneqsc}) is promising, where wide scaling region results in high precision 
in comparison with a simple power-law fitting of equilibrium data~\cite{Yasuda13}, 
though accuracy of the estimation is still not comparable to that of a detailed study 
on critical phenonena in the 3D classical Heisenberg model~\cite{Campostrini02}. 
We made a similar analysis on the staggered susceptibility, and obtained the critical 
exponent $\gamma/\nu$ consistent with the hyperscaling relation~\cite{suscexp}.

\smallskip
\par
{\it Summary and discussion}. 
In the present article we generalized the cluster nonequilibrium 
relaxation (NER) scheme to quantum phase transitions. Since modern quantum 
Monte Carlo algorithms such as the loop algorithm are based on cluster updates, 
the present scheme is indispensable for the NER analysis of quantum 
systems. As an example, we considered the N\'eel-dimer quantum 
phase transition in the two-dimensional $S=1/2$ columnar-dimerized 
antiferromagnetic Heisenberg model on a square lattice. This model 
is the simplest isotropic quantum spin system to exhibit a quantum 
phase transition with respect to the strength of dimerization $\delta$, 
and belongs to the universality class of the 3D classical Heisenberg model.

In the present study, numerical calculations were 
based on the continuous-time loop algorithm with the 
Ising basis and started from the isolated dimer state. 
Although we have numerically and theoretically clarified that physical quantities 
at the critical point show the stretched-exponential relaxation behavior in the 
early-time relaxation in cluster algorithms in classical spin systems, this behavior 
is not trivial in the present case, because one-dimensional loop clusters are 
geometrically different from the bulky ones in the Swendsen-Wang and Wolff algorithms. 
Then, the critical point $\delta_{\rm c}$, 
critical exponent $\beta/\nu$ and relaxation exponent $\sigma$ were estimated 
from the nonequilibrium-to-equilibrium scaling plot of the absolute value of the 
N\'eel order as in our previous study on the 3D classical Heisenberg model. 
The present estimates $\delta_{\rm c}=0.90947(5)$, $\beta/\nu=0.514(2)$ 
and $\sigma=0.502(9)$ are comparable to previous studies 
(even more precise than previous equilibrium QMC simulations in $\beta/\nu$). 
Consistency with the 3D classical Heisenberg model holds not only 
for $\beta/\nu$ but also for $\sigma$. These results reveal that the cluster NER 
scheme can be generalized to quantum phase transitions on the basis of the 
continuous-time loop algorithm.

As shown in the present article, large numbers of random-number 
sequences should be averaged for accurate data in the NER analysis 
instead of long-time measurements in the equilibrium Monte Carlo analysis, 
and such averaging can be replaced by sample averaging in random systems. 
This fact indicates that numerical efforts in random systems may not be so 
different from pure systems, which is the essential merit of the present scheme. 
Studies along this direction is now in progress. 
\smallskip
\par
{\it Acknowledgments}. 
Y.~N.\ thanks K.~Harada for helpful comments. 
The present study was supported by JSPS KAKENHI Grant No.~16K05493. 
The random-number generator MT19937~\cite{MT} was used for 
numerical calculations. Most calculations were  performed on the 
Numerical Materials Simulator at National Institute for Materials Science.


\begin{thebibliography}{99}
\bibitem{Trotter59}
H.~F.~Trotter, Proc.\ Am.\ Math.\ Soc.\ {\bf 10}, 545 (1959).
\bibitem{Suzuki76}
M.~Suzuki, Prog.\ Theor. Phys.\ {\bf 56}, 1454 (1976).
\bibitem{Beard96}
B.~B.~Beard and U.-J.~Wiese, Phys.\ Rev.\ Lett.\ {\bf 77}, 5130 (1996).
\bibitem{Evertz93}
H.~G.~Evertz, G.~Lana, and M.~Marcu, 
Phys.\ Rev.\ Lett.\ {\bf 70}, 875 (1993).
\bibitem{Prokovev98}
N.~V.~Prokov'ev, B.~V.~Svistunov, and I.~S.~Tupitsyn, 
Sov.\ Phys.\ JETP {\bf 87}, 310 (1998).
\bibitem{Syljuasen02}
O.~F.~Sylju{\aa}sen and A.~W.~Sandvik, Phys.\ Rev.\ E {\bf 66}, 046701 (2002).
\bibitem{NERrev}
As a review on the NER method, 
Y.~Ozeki and N.~Ito, J.\ Phys.\ A: Math.\ Theor.\  {\bf 40}, R149 (2007).
\bibitem{SW}
R.~H.~Swendsen and J.-S.~Wang, Phys.\ Rev.\ Lett.\ {\bf 58}, 86 (1987).
\bibitem{Wolff}
U.~Wolff, Phys.\ Rev.\ Lett.\ {\bf 62}, 361 (1989), 
Nucl.\ Phys.\ B {\bf 322}, 759 (1989).
\bibitem{Berg}
B.~A.~Berg and T.~Neuhaus, Phys.\ Rev.\ Lett.\ {\bf 68}, 9 (1992).
\bibitem{Hukushima}
K.~Hukushima and Y.~Nemoto, J.\ Phys.\ Soc.\ Jpn.\ {\bf 65}, 1604 (1996).
\bibitem{WL}
F.~Wang and D.~P.~Landau, Phys.\ Rev.\ Lett.\ {\bf 86}, 2050 (2001).
\bibitem{Nonomura98}
Y.~Nonomura, J.\ Phys.\ Soc.\ Jpn.\ {\bf 67}, 5 (1998), 
J.\ Phys.\ A: Math.\ Gen.\ {\bf 31}, 7939 (1998).
\bibitem{Nakamura03}
T.~Nakamura and Y.~Ito, J.\ Phys.\ Soc.\ Jpn.\ {\bf 72}, 2405 (2003).
\bibitem{Nonomura14}
Y.~Nonomura, J.\ Phys.\ Soc.\ Jpn.\ {\bf 83}, 113001 (2014).
\bibitem{Nonomura15}
Y.~Nonomura and Y.~Tomita, Phys.\ Rev.\ E {\bf 92}, 062121 (2015).
\bibitem{Nonomura16}
Y.~Nonomura and Y.~Tomita, Phys.\ Rev.\ E {\bf 93}, 012101 (2016).
\bibitem{Tomita18}
Y.~Tomita and Y.~Nonomura, Phys.\ Rev.\ E {\bf 98}, 052110 (2018).
\bibitem{Matsumoto01}
M.~Matsumoto, C.~Yasuda, S.~Todo, and H.~Takayama, 
Phys.\ Rev.\ B {\bf 65}, 014407 (2001).
\bibitem{Wenzel09}
S.~Wenzel and W.~Janke, Phys.\ Rev.\ B {\bf 79}, 014410 (2009).
\bibitem{Yasuda13}
S.~Yasuda and S.~Todo, Phys.\ Rev.\ E {\bf 88}, 061301(R) (2013).
\bibitem{QMCrev}
As a review on the loop algorithm, 
N.~Kawashima and K.~Harada, J.\ Phys.\ Soc.\ Jpn.\ {\bf 73}, 1379 (2004).
\bibitem{comment}
During the nonequilibrium relaxation process from the 
isolated dimer state, the $(2,1)$-Binder ratio 
$B_{{2,1}}(t,L)  \equiv \langle m_{\rm N}^{2}(t,L) \rangle / \langle |m_{\rm N}(t,L)| \rangle^{{2}}$ 
shows a nontrivial peak after several MCS, even though it is expected 
to decrease monotonically from the Gaussian value $\pi/2$. 
\bibitem{Campostrini02}
M.~Campostrini, M.~Hasenbusch, A.~Pelissetto, P.~Rossi, 
and E.~Vicari, Phys.\ Rev.\ B {\bf 65}, 144520 (2002).
\bibitem{suscexp}
The critical exponent $\gamma/\nu$ of the staggered susceptibility 
$\chi_{\rm st} \equiv \frac{1}{N}\sum_{i,j} (-1)^{i-j} S_{i}^{z}S_{j}^{z}$ can 
be evaluated similarly, and we obtain $\gamma/\nu = 1.973(4)$. 
This estimate satisfies $2 \beta/\nu + \gamma/\nu = 3.001(6)$, 
which is consistent with the hyperscaling relation, 
$2\beta/\nu + \gamma/\nu = d+1$ with the spatial dimension $d=2$ 
plus the imaginary-time dimension.
\bibitem{MT}
M.~Matsumoto and T.~Nishimura, ACM TOMACS {\bf 8}, 3 (1998). 
Further information is available from the Mersenne Twister 
Home Page, currently maintained by M.~Matsumoto.
\end{thebibliography}
\end{document}